\documentstyle[psfig,12pt]{article}
\title{\bf {Formation Scenarios }}
\author{Rosemary F.~G.~Wyse\\
\vspace{1cm}\\
\normalsize Department of Physics \& Astronomy, The Johns Hopkins 
University,\\ 
\normalsize Baltimore, MD 21218, USA\\}
\date{\mbox{}}
\begin{document}
\maketitle
\pagestyle{empty}
%
%
\def\bull{\vrule height .9ex width .8ex depth -.1ex}
\makeatletter
\def\ps@plain{\let\@mkboth\gobbletwo
\def\@oddhead{}\def\@oddfoot{\hfil\tiny\bull\quad
``The Galactic Halo : from Globular
Clusters to Field Stars'';
35$^{\mbox{\rm th}}$ Li\`ege\ Int.\ Astroph.\ Coll., 1999\quad\bull}%
\def\@evenhead{}\let\@evenfoot\@oddfoot}
\makeatother
%
%
\def\beginrefer{\section*{References}%
\begin{quotation}\mbox{}\par}
\def\refer#1\par{{\setlength{\parindent}{-\leftmargin}\indent#1\par}}
\def\endrefer{\end{quotation}}
\def\etal{{\it et al.\/}}
%
%
{\noindent\small{\bf Abstract:} I discuss various proposed formation
scenarios for the metal-poor components of the Milky Way Galaxy,
emphasising the stellar halo and the thick disk.  Interactions and
accretion played a significant role in Galactic evolution, in
particular at earlier epochs.  The present observations favour a
scenario by which the thick disk formed through the heating of a
pre-existing thin stellar disk, with the heating mechanism being the
merging of a satellite galaxy. A remnant `moving group' of the
satellite would provide strong support for this scenario, and may have
been detected.  The field stars in the stellar halo probably formed in
early small-scale star-forming regions, which subsequently disrupted.
Late accretion is not important for the bulk of the stellar halo.  The
stellar initial mass function shows no evidence of variations, and
indeed shows evidence of being invariant, even in companion satellite
galaxies.}

%
%
\section{Introduction}

In keeping with the focus of the Colloquium, I will discuss primarily the
formation of the thick disk and of the stellar halo, the more
metal-poor components of the Milky Way Galaxy.  However, we should
bear in mind that at approximately the same epoch at which these stars
formed, around 12 or 14~Gyr ago, the stars in the central bulge were
also forming (Ortolani \etal\ 1993; Feltzing \& Gilmore 1999), despite
the bulge stars being significantly more metal-rich in the mean than
are stars in either the thick disk or the stellar halo.  Further,
there also exist old stars in the local thin disk, at least as old as
10~Gyr and perhaps as old as stars in the halo (e.g.~Edvardsson 
\etal\ 1993), which could imply that
there was no significant hiatus between halo and the onset of disk
formation (we have little information for the age distribution of
stars in the more distant thin disk, either interior or exterior to
the solar circle).  The existence of these old disk stars poses
significant problems for some models of disk galaxy formation, such as
those (Weil, Eke \& Efstathiou 1998) that posit the delay of disk
formation until after a redshift of unity, or lookback times of only $\sim 
7$~Gyr for a flat matter-dominated Universe with Hubble constant of
65~km/s/Mpc.  

Current structure formation scenarios favour hierarchical clustering,
such as that in a universe dominated by cold, dark matter (CDM).  In
this picture, the first objects to turn around and collapse have a
mass that is a small fraction of the mass of a galaxy like the Milky
Way (e.g.~Tegmark \etal\ 1997), and larger-scale structure grows by
clustering and merging of this small-scale structure. The dynamics of
the dark matter, interacting only through gravity, is rather
straightforward to model, through N-body simulations and semi-analytic
techniques such as the Press-Schechter formalism (e.g.~Lacey \& Cole
1993). Following the behaviour of the baryons and predicting the
evolution of luminous galaxies is much more difficult, either with
gas-dynamic simulations (e.g.~Navarro \& Steinmetz 1997; Pearce \etal\
1999) or star-formation prescriptions combined with the N-body
simulations and/or semi-analytic treatment of the merging of the dark
haloes (e.g.~Kauffmann \etal\ 1999; Baugh \etal\ 1998).  Cold dark
matter models have found much success in the analysis of observations
of large-scale structure, from the microwave background down to
clusters of galaxies.  Further, there are many examples in both the
local, and distant, Universe of interacting and merging galaxies.  The
Milky Way is clearly interacting with its satellite galaxies, such as
the LMC/SMC (Putman \etal\ 1998) and the Sagittarius dwarf spheroidal
galaxy (Ibata, Gilmore \& Irwin 1994; 1995).

However, disk galaxies as observed, with a broad range of stellar ages
in the thin disk, cannot have experienced merger events that were too
frequent or too violent, since this would have destroyed the disk
(cf.~Ostriker 1990; Toth \& Ostriker 1992). The past merging with
dissipationless stellar or dark-matter systems is restricted to the
accretion of small objects onto a dominant central system.  And the
accreted objects have to be assimilated quite efficiently, since at
least in the Milky Way there is little evidence of successive,
significant, past mergers.

This last point is particularly difficult in light of new
high-resolution N-body simulations by two groups (Klypin \etal\ 1999;
Moore \etal\ 1999a) which for the first time have enough dynamic range
to model both large and small scales within the same simulation.  These
simulations have been restricted so far to flat ($\Omega=1$)
cosmologies, with CDM the dominant mass, and include a universe
dominated by $\Lambda$ at the present day ($\Omega_\Lambda =
0.7$, $\Omega_{CMD} = 1 - \Omega_\Lambda$), as favored by a variety of
constraints on large scales (Bahcall \etal\ 1999).  In these model universes,
small-scale dark haloes are very persistent (essentially reflecting
their high redshifts of formation and hence high density) and they
predict that a galaxy like the Milky Way should today have around a
factor of ten more satellite galaxies than we observe. Of course, the
simulations are strictly restricted to dark haloes only, and one might
postulate that the `missing' satellites (cf. Klypin \etal\ 1999) are
dark, or perhaps related to the extremely high-velocity clouds
identified by Blitz \etal\ (1999). However, even if dark, those
satellites on radial orbits that interact  with the disk of the Milky Way 
could make a {\it thin\/} disk impossible to sustain (Moore
\etal\ 1999a).

Allowing an open universe, still dominated by CDM, would change the
timescales of the growth of structure, but not remove the basic
problem with the over-prediction of the number of long-lived satellite
galaxies. The simulations of a flat universe 
provide very good agreement with observations on larger scales, 
such as the galaxy
luminosity function within clusters of galaxies (Moore \etal\ 1999a),
leading to the suggestion that some modification be made to the CDM
power spectrum, to reduce small-scale power.  Such a modification, truncating the power spectrum at small scales, may 
also be favoured by the discrepancies between the shapes of the
galaxy rotation curves predicted by standard 
CDM-dominated models (Navarro, Frenk \&
White 1997) and the observations for dwarf galaxies (Moore 1994) and
for other apparently dark-matter-dominated galaxies, such as
low-surface-brightness disks (Burkert \& Silk 1997; 1999; Moore \etal\
1999b).

Indeed, many of the properties of present-day disks, for example their
scale-lengths and rotational velocities, can be explained by simple
dissipational collapse of gas within a fixed dark halo potential, with
detailed conservation of angular momentum (Fall \& Efstathiou 1980;
Dalcanton \etal\ 1997; Mo, Mao \& White 1998), followed by modest
re-arrangment by, for example, viscous processes (e.g.~Zhang \& Wyse
2000). Merging of dark haloes and luminous galaxies is usually
accompanied by angular-momentum transport, driven by gravitational
torques and dynamical friction, removing angular momentum from the
disk material, and `standard' CDM-dominated models produce disks that
are too small (Navarro \& Steinmetz 1997). Again, this suggests only
`minor-mergers' in the history of a disk galaxy. Of course the
formation of stars is set by local gravitational instability, on the
scale of giant molecular clouds and their cores. 

Thus essentially all models of galaxy formation and evolution invoke
early star formation in substructure, whether the smaller scales
formed by fragmentation (perhaps reflecting the Jeans mass) or were
primordial density fluctuations (such as the CDM power spectrum).
Questions that both observation and theory should address include what
fraction of this substructure survives to the present, what are the
possible relationships of this early substructure with present-day
smaller systems, e.g.~dwarf galaxies or globular clusters, and how
important are interactions with satellite galaxies?  The faint stellar
Initial Mass Function (IMF) in satellite galaxies of the Milky Way,
which are possible examples of surviving `building blocks', is also
now accessible to direct determination.  Local observations of old
stars provide complementary constraints to observations of high
redshift objects.

\section{The Thick Disk}
This stellar component was first detected in the Milky Way Galaxy
through star counts (Gilmore \& Reid 1983), although surface
photometry of external S0 galaxies had earlier revealed `thick disks'
in them (Burstein 1979; Tsikoudi 1979).  Perhaps the most recent study
of the structure of the thick disk through star counts is that of
Phleps \etal\ (1999), who utilised the R-band data in the NGP of the
Calar Altar (faint galaxy) survey; they detect the thick disk, and
derive parameter values that are reasonably consistent with earlier
results, in that the scale-height of the thick disk is around 1kpc,
and the local (solar neighbourhood) normalisation by number is several
percent.

\subsection{Connection to Other Galactic Components}
At the time of its discovery, it was postulated that the thick disk
was formed by local compression of the stellar halo by the potential
of the thin disk (Gilmore \& Reid 1983).  This was soon disproven
(Gilmore \& Wyse 1985) by the determination of distinct metallicity
distributions of thick disk and stellar halo (see Fig.~1 here).

\begin{figure}[!htb]
\vskip -0.75in
\hskip 0.75in
\psfig{file=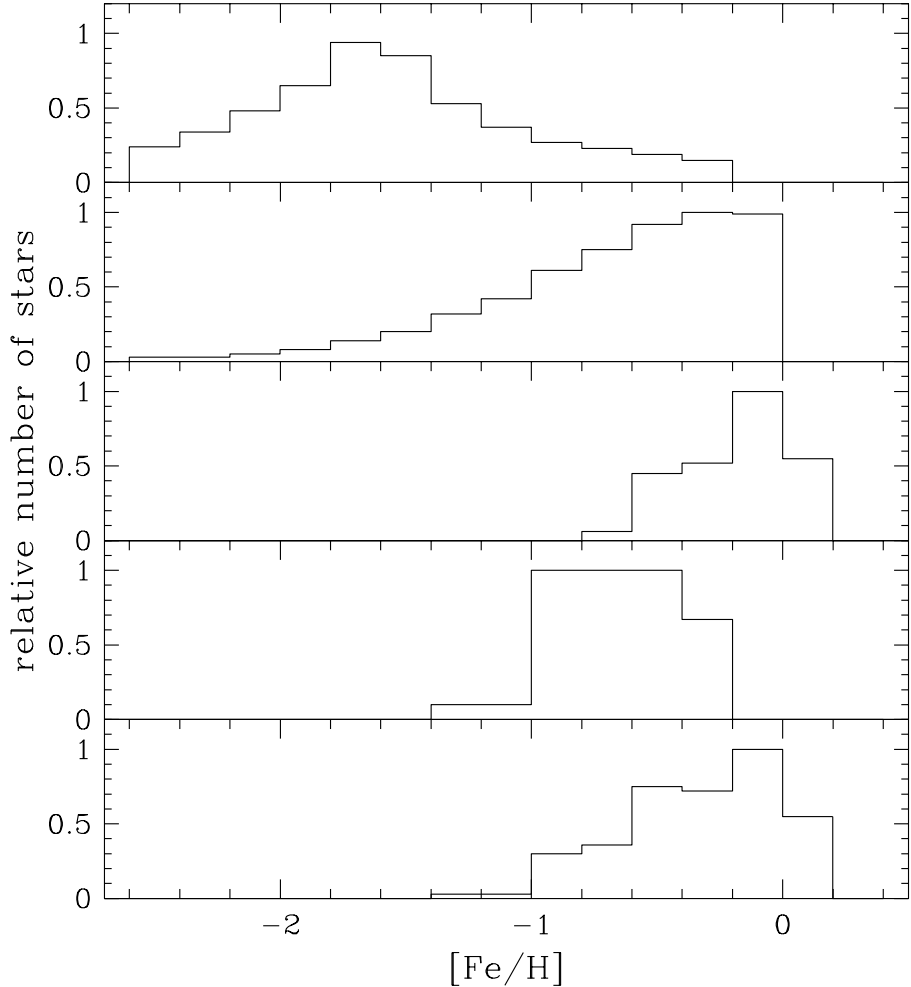,width=5.0in}
\caption{The metallicity distributions of representatives of the stellar
populations of the Milky Way Galaxy.  Where possible, a measure of the
true iron abundance is plotted.  The panels are (top to bottom) the
local stellar halo (Carney \etal\ 1994, their kinematically-selected
sample); the outer bulge K-giants (Ibata \& Gilmore 1995), truncated at solar metallicity due to calibration limitations; the
volume-complete local thin disk F/G stars (derived from the
combination of the Gliese catalogue and {\it in situ\/} survey); the
volume-complete local thick disk F/G stars (derived similarly); and 
lastly the `solar
cylinder', i.e. F/G stars integrated vertically from the disk plane to
infinity. This figure is based on Fig.~16 of Wyse \& Gilmore (1995).}
\end{figure}

But is the thick disk simply the extreme thin disk, meaning they have
a common origin? An increase in scale-height and velocity dispersion
with stellar age within the thin disk is well-established (e.g.~Wielen
1977).  However, the thick disk is discontinuous in its kinematics
from the thin disk (Wyse \& Gilmore 1986; Gilmore, Wyse \& Kuijken
1989).  Further, the value of the vertical velocity dispersion of the
thick disk, some $40-45$km/s (e.g.~the review of Majewski 1993 and
references therein) is too high to be explained by the known heating
mechanisms for the stars in the thin disk, namely interactions with
local gravitational perturbations in the disk, such as Giant Molecular
Clouds (Spitzer \& Schwartzschild 1957; Lacey 1991).  More exotic
phenomena, such as close encounters with massive black holes in the
dark halo, can provide the required high amplitude of random motions
for a small fraction of the thin-disk stars (Ostriker \& Lacey 1985;
see also Sanchez-Salcedo 1999), but then the thick disk should be a
random sample of the thin disk, and have very a similar stellar
population.  Again, the different metallicity distributions of thick
disk and thin disk argue against this -- the metallicity distribution
of the thick disk peaks at [Fe/H]$ \sim -0.6$~dex, and is rather broad
(Gilmore \& Wyse 1985; Carney, Latham \& Laird 1989; Wyse \& Gilmore
1995; Bonifacio \etal\ 1999), while that of the thin disk peaks around
$-0.2$~dex (e.g.~Wyse \& Gilmore 1995; see Fig.~1 above).  Further,
the thick disk is apparently composed of only old stars, ages older
than $\sim 12$~Gyr (Gilmore \& Wyse 1985; Gilmore, Wyse \& Kuijken
1989; Carney, Latham \& Laird 1989; Edvardsson \etal\ 1993; Gilmore,
Wyse \& Jones 1995; Fuhrmann 1998; Carney, this volume; Bartasiute \&
Lazauskaite, this volume).  The age distribution older than this is
not well-defined with present data, but a younger component must be
only minor. This last point is illustrated in Fig.~2, and argues
strongly against an extended period of thick disk formation, and in
favour of a unique event, long ago.
\begin{figure}[!htb]
\hskip 0.75in
\psfig{file=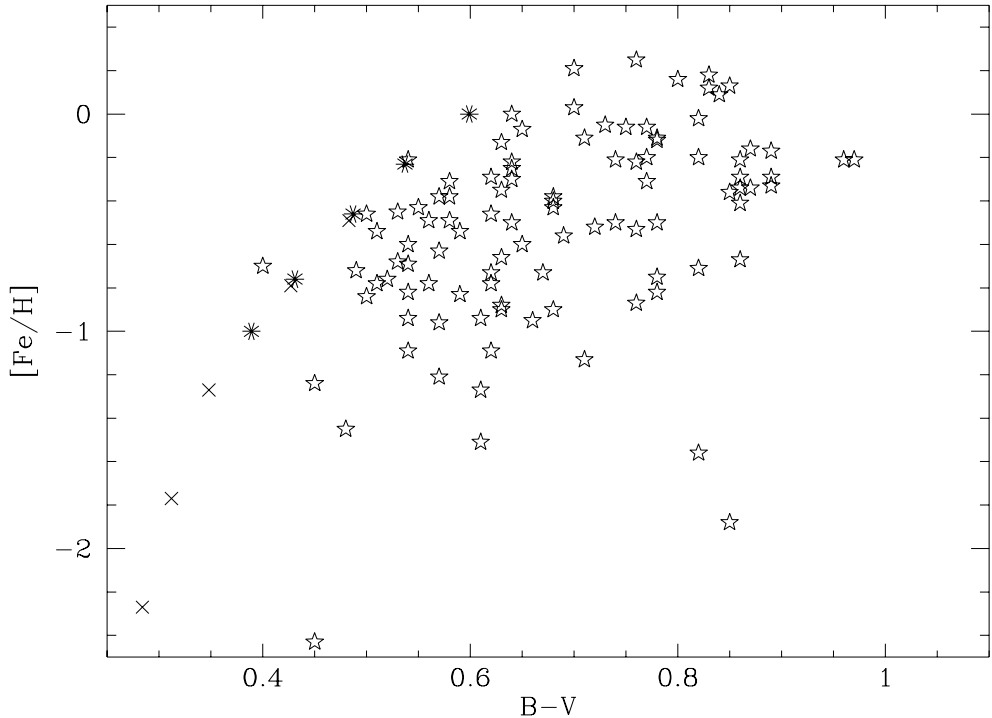,width=5.5in,angle=270}
\caption{Scatter plot of iron abundance {\it vs\/} B$-$V colour for
thick disk F/G stars, selected {\it in situ\/} at 1-2kpc above the Galactic
Plane (stars), together with the 12~Gyr turnoff colours (crosses) from
VandenBerg \& Bell (1985) and from VandenBerg (1985; asterisks).  The vast
majority of thick disk stars lie to the red of the 12~Gyr turnoff
points, indicating that few, if any, stars in this population are younger than
this age.  This figure is based on Fig.~6 of Gilmore, Wyse \& Jones (1995).}
\end{figure}

\subsection{Heating or Cooling?}
Two possibilities remain, one that the thick disk formed as part of
the (dissipational) settling of the proto-thin disk, the second that
the thick disk formed during a traumatic heating event early in the
evolution of the thin disk.  In the former (cooling) 
scenario the scaleheight of
the stellar disk decreases with time, and is set by a balance between
cooling (and star formation) and gravity; the discontinuity between
thick and thin disks could reflect the change in the cooling law as
metallicity increases above $\sim -1$~dex and line radiation from
metals becomes dominant (Gilmore \& Wyse 1985; Burkert, Truran \& Hensler 
1993; Burkert \& Yoshii 1996).  One might then expect all
(moderately metal-rich) disk galaxies to have a thick disk, which they
do not (e.g.~Shaw \& Gilmore 1990; Fry \etal\ 1999).

The latter (heating) scenario draws some support from the fact that,
as noted above, interactions between the Milky Way and its satellites
are ongoing.  The vertical velocity dispersion of the thick disk can
be provided for if a significant part of the orbital energy of a
moderate-mass satellite galaxy is transformed into additional internal
energy of the stellar thin disk (Gilmore \& Wyse 1985; Ostriker 1990;
Majewski 1993).  The effect of the accretion of a companion galaxy on
the disk depends on many parameters such as those of the satellite's
orbit (initial inclination to the disk plane, pericenter and apocenter
distances, sense of angular momentum) and the satellite's density
profile and total mass.  Simulations of the merging process between a
stellar disk and satellite have become increasingly sophisticated in
recent years, including more physics such as allowing the excitation
of the internal degrees of freedom of the dark halo, which lessens the
heating effect on the disk (e.g.~Huang \& Carlberg 1997; Walker, Mihos
\& Hernquist 1996; Velazquez \& White 1999).  The extant simulations
suggest that the accretion by the present-day stellar disk of a
stellar satellite with mass some 20\% of that of the disk can produce
a thick disk similar to that observed in the Milky Way.  However, gas
has yet to be included in the simulations investigating disk
heating, which is an important shortcoming, since gas if present
(which is likely), would absorb and subsequently radiate away some of
the orbital energy, again lessening the impact of the merger.

Further, the initial conditions of the published simulations assume a
fully-assembled stellar disk, and especially given the results above
on the old age of the Galactic thick disk, we need simulations that
better model conditions at an early stage of disk galaxy evolution.
The stellar population in the local thin disk is consistent with a
roughly constant star-formation rate over a Hubble time
(e.g.~Rocha-Pinto \& Maciel 1997), implying that the local stellar
disk at the lookback time corresponding to the formation of the thick
disk was around 10\% of its present mass.  This is close to the mass
of the local thick disk, expressed as a fraction by mass of the
present-day local thin disk, suggesting essentially all of the
pre-existing thin disk was heated.  Of course one needs to know how to
generalise this result to the entire disk, which requires global
knowledge of star-formation histories and thin- and thick-disk
structural parameters.  Further, the energy losses due to gas have yet
to taken into account.  Really one needs a cosmologically
self-consistent model including disk buildup, with appropriate star
formation.

As a corollary to this heating scenario, if a significantly massive
satellite were responsible for the thick disk, the accompanying
torques could drive a substantial fraction of the gas in the disk at
that time to the central regions, perhaps triggering rapid star
formation (cf.~Hernquist \& Mihos 1995) and even  the formation of
the `bulge' (Minniti 1996) or `thick disk' (Armandroff 1989) globular
clusters.  It may well be no coincidence that the ages of field stars
in the bulge (e.g.~Feltzing \& Gilmore 1999) and in the thick disk are
similar.

The lack of young or even intermediate-age stars in the thick disk
limits the last significant merger to have occurred a long time ago,
at lookback times greater than around 12~Gyr, or redshifts of $\sim 6$
in standard matter-dominated flat cosmology. This is rather difficult
for $\Omega_{CDM}=1$ models, requiring the Milky Way to be an unusual
galaxy (cf.~Toth \& Ostriker 1992).  Note that Moore \etal\ (1999a)
argue that any hierarchical clustering cosmology, even the open
CDM-dominated model, has problems since not all disk galaxies have
thick disks.  However, I feel that given the many parameters
determining the effect of a merger on a disk, a widely disparate
population of thick disks, including one too small to have been
detected, must result.

\subsection{Where is the Shredded Satellite?}
Is there then a signature of the remnant of the putative satellite
that caused the Milky Way thick disk? Stars that are removed by tides
will remain on orbits close to that of the centre-of-mass of the
satellite at the time the stars are removed (e.g.~Tremaine 1993;
Johnston 1998); the orbit of the satellite is expected to decay
through dynamical friction (at a rate dependent on its mass, but the
effect should be significant for the $\sim 20\%$ fractional masses
under consideration), depositing `shredded satellite' stars over a
reasonably large spatial region.  These stars will phase mix.
Published simulations of low-inclination satellite orbits indeed show
that the satellite is dispersed into a broadly flattened distribution,
mixed in with the heated thin-disk stars.  Satellites on prograde
(rather than retrograde) orbits couple better to the disk and provide
more heating, and are favoured to cause the thick disk (e.g.~Velazquez
\& White 1999). Thus one might expect a signature to be visible in the
mean orbital rotational velocity of the stars, and for a typical satellite
orbit in the mean the stripped stars would 
lag the Sun by more than does the canonical thick disk.
The relative number of stars in the `shredded satellite' versus the
heated-thin disk (now the thick disk) depends on the details of the
shredding and heating processes, and is a diagnostic of them, and may
well vary strongly with location.  Note that a merger without
accompanying heating of the disk, but producing the thick disk in its
entirety from the shredded satellite is rather contrived: this would
require a fairly massive satellite, given what we know of the
mass-metallicity relationship and the relatively high mean metallicity
of the thick disk -- which at $\sim -0.6$~dex is greater than that of
the old population in the LMC, and about equal to the young population
in the SMC (see, e.g.~de Freitas Pacheo, Barbuy \& Idiart 1998) -- and
also the total luminosity of the thick disk, extrapolated from its
locally-defined structural parameters (several percent of the present
thin disk).  Only a very narrow part of parameter space could allow a
satellite this massive to penetrate even to the solar circle, then be
tidally disrupted without imparting any damage to the thin disk.

We (Gilmore, Wyse, Norris \& Freeman) are quantifying the 
phase-space structure of 
the Milky Way, through a comprehensive statistical study of the
kinematics and metallicity distributions of stars in the interface
between the thick disk and stellar halo, those stellar components for
which mergers are most often implicated.  We are
using the 2-degree-field fibre-fed spectrograph on the
Anglo-Australian telescope, providing 400 spectra simultaneously. These 
spectra  are used to obtain radial velocities and
absorption line-strengths for samples of F/G main sequence stars at
distances from the Sun of 5--10kpc (dependent on metallicity and
magnitude), beyond significant contamination by the thin disk, down
several key lines-of-sight.  Our targets include fields towards and
against Galactic rotation, to provide optimal halo/thick disk
discrimination through orbital angular momentum. Further, we have
fields at the same Galactic longitude but different latitudes, to
determine if any kinematic features are halo-like or disk-like in
their spatial distributions.  We also include lines-of-sight interior
to and exterior to the Sun, to allow characterisation of the velocity
dispersion tensor, detection of gradients and any
metallicity--kinematics correlation, and crucially to test if
kinematic structure is restricted to angular momentum, or is equally
present in the other components of the velocity tensor.  Our approach
investigates the time-integrated structure of the halo and thick disk
and is quite complementary to surveys of the far outer Galaxy.

Our first observations have apparently detected a new kinematic
component of the Milky Way Galaxy, plausibly the shredded remnant of
the satellite whose merger with the Galaxy produced the canonical
thick disk, by heating the pre-existing thin disk.  As described
above, depending on the mass, density profile and orbit of the
satellite, `shredded-satellite' stars will leave a kinematic
signature, distinct from the canonical thick disk that results from
the heated thin disk.  As shown in Fig.~3 and Fig.~4 below, the radial
velocity distributions for our samples indeed show evidence for an
excess number of stars moving on orbits with V-velocity between those
of the canonical thick disk and halo, with a low velocity dispersion,
plausibly smaller than either of these Galactic components, and
further this signature is strongest in the blue stars.

A similar kinematic 
signature was seen in our earlier {\it in situ\/} (AUTOFIB)
survey of the kinematics and metallicity distributions of stars in the
thin disk/thick disk interface (Wyse \& Gilmore 1990).  Further, Fuchs
\etal\ (1999) have reported a similar result, finding an excess number
of stars with intermediate kinematics in a narrow metallicity range,
for a local sample, based on the kinematically-selected sample of
Carney \etal\ (1996), supplemented with Hipparcos data.  These stars
are best interpreted as being the actual debris of the satellite and
would provide the `smoking gun' signature of the most significant
merger in the Galaxy's past.  They must be part of a fairly large
system, being detected in disparate samples, but this requires a
larger statistically-significant sample for confirmation, to allow
quantification of the properties of the satellite galaxy, and to
distinguish between this interpretation and a vertical gradient in the
mean rotation of the thick disk (e.g.~Majewski 1993; Mendez \etal\
1999).  The metallicity distribution will be an important constraint,
reflecting that of the parent satellite in one case (cf.~Freeman 1993), 
and that intrinsic to the thick disk in the other (that of the old thin disk,
should the canonical thick disk be the heated thin disk).

\begin{figure}[!htb]
\vskip -2.2in
\psfig{file=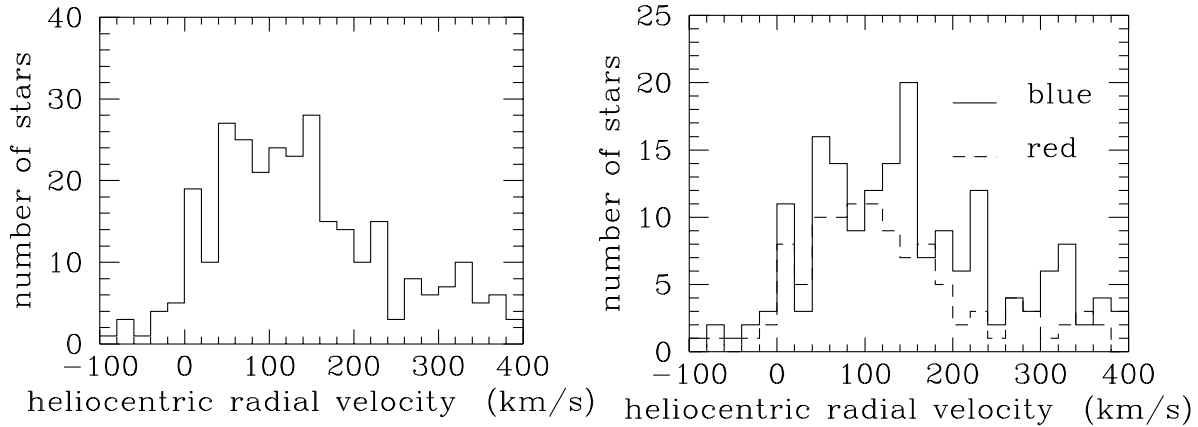,width=6.5in,angle=270}
\caption{Histograms of the heliocentric
radial velocity for the field at ($\ell = 270$, $b = +33$); this
line-of-sight was chosen to probe the orbital angular momentum of the
stars, with $v_{radial} = -0.84 V + 0.54W$.  The left panel shows all
the stars from one observing run, while the right panel shows them divided by
colour. The standard, slightly-prograde, stellar halo would have a
fairly broad radial velocity distribution peaking at around 200km/s
and dispersion of 100km/s (see Fig.~4 below), and the thick disk
peaking at around 25km/s, with dispersion of 55km/s.  
There is an apparent excess of stars with line-of-sight velocity 
intermediate to these peaks, and a poorly-defined but low dispersion, 
and the signature is stronger for the bluer
stars.}
\end{figure}
\begin{figure}[!htb]
\vskip -2in
\hskip 0.2in
\psfig{file=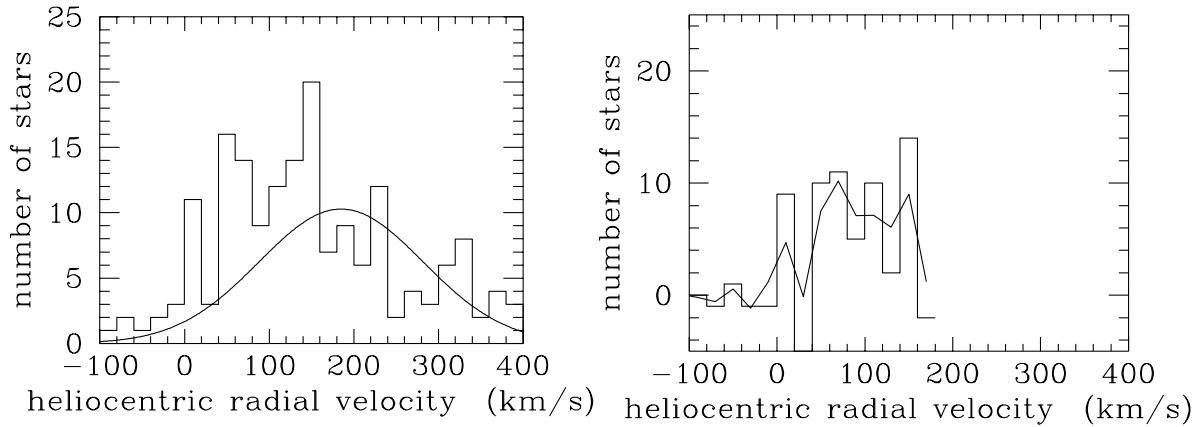,width=6.5in,angle=270}
\caption{The left panel is all the blue stars from Fig.~3, with the
predictions in this line-of-sight for a standard, slightly-prograde,
halo.  The right panel is the `excess population' (plus a running
mean) obtained by reflecting the data about the mean of this symmetric
distribution function.}
\end{figure}
\section{The Stellar Halo}

To first order, the stellar halo is the metal-poor, old,
slowly-rotating, extended but centrally-concentrated, stellar system
represented at the solar neighbourhood by the high-velocity subdwarfs.
The stellar halo is usually distinguished from the central bulge when
discussing the Milky Way, but it is common practice simply to refer to
them as one entity -- the bulge or spheroid -- in discussions of
external disk galaxies (see Wyse, Gilmore \& Franx 1997 for a review).
This is of course related to the difficulty of detecting a
component like the stellar halo in external galaxies -- the mass ratio
between bulge and stellar halo in the Milky Way is around a factor of
ten, and the surface brightness of the stellar halo at the solar
circle is around 28 mag/sq arcsec in the B-band (Morrison 1993).

Essentially all models of the formation and evolution of the stellar
halo invoke early star formation in small substructure, with
subsequent disruption of these systems and mixing and assimilation of
the stars formed therein into the field stellar halo.  The small
substructure may have formed through fragmentation of an initially
`monolithic' baryonic component, perhaps reflecting the Jeans mass of
shock-heated gas in the potential well of a larger-scale protogalaxy
(e.g.~Fall \& Rees 1985), or could reflect simply the initial power
spectrum of primordial density fluctuations (e.g.~White \& Rees 1978).
Or, a combination of the above.  It is highly likely that the stellar
halo is `multi-component' and we need to quantify the stellar masses
in the different components, and characterise their origins.
 
Questions that can be addressed and hopefully answered by the fossil
record of the halo stars include -- when did the stars form?    
Did internal (e.g.~feedback from massive
stars) or external (e.g.~collisions or tidal disruption) processes
cause the destruction of the substructure?  When was
the substructure disrupted? What is the relation, if
any, between this putative substructure and surviving systems in the halo
such as globular clusters and satellite galaxies?  What is the
relationship to `Population III', those stars formed very early, at
redshifts prior to re-heating and re-ionization of the InterGalactic Medium?
Has accretion of stars and stellar systems played an important role?

\subsection{Elemental Abundances}
The chemical elemental abundances of a typical halo star are as
expected if only Type~II supernovae enriched the gas out of which the
stars formed (e.g.~Nissen \etal\ 1994; Norris 1999).  The most
straightforward explanation of this observation is that these stars
formed in star-formation events that were of short duration, shorter
than the time needed to have significant production of
newly-synthesised material by Type~Ia supernovae.  While there is not
yet a generally-accepted model of Type~Ia supernovae (other than
involving a white dwarf driven over the Chandrasekhar mass limit via
accretion of some sort) several scenarios predict timescales after
formation of the progenitor main sequence stars, for significant
chemical enrichment by Type Ia supernovae, of around 1~Gyr
(e.g.~Smecker \& Wyse 1991; Yungelson \& Livio 1998).  One does not
require that the entire stellar halo formed on this timescale, only
that self-enriching regions formed stars this rapidly, and there was
little cross-contamination between non-synchronized regions.  Indeed,
an attractive mechanism to produce only a short duration to star
formation is a Type~II supernovae-driven wind, more naturally-produced
if the star-forming regions have local potential wells significantly
shallower than does the halo as a whole.  

Thus one is led to a picture
wherein the field stars of the halo form in fragile fragments/blobs,
within which feedback from massive stars can be sufficiently
disruptive that star formation is truncated, and a large part of the
remaining interstellar medium ejected.  The feedback and mass loss
could be sufficient to unbind the `fragment' totally, or it could be
that the new virial equilibrium of the `fragment' is sufficiently
fragile that external processes such as tidal forces can disrupt the
`fragment'.  The ejected gas will cool and dissipate, 
and with angular
momentum conservation will settle into the central regions of the
overall larger-scale potential well of the proto-Galaxy, somewhat as
envisaged by Eggen, Lynden-Bell \& Sandage (1962). This could form the
central bulge.
As pointed out by Hartwick (1976), one can understand the low
mean metallicity of the halo within models with a fixed stellar IMF if
there is gas removal during the formation of the halo stars -- 
a reduction of a factor of
around 10 in the mean metallicity, compared to theoretical expectations
with no gas loss, is required for the stellar halo, and this is
achieved by removing around this ratio of mass during star formation.
Thus one would predict a central bulge some 10 times more massive than
the stellar halo, in agreement with estimates of the masses of stellar 
halo and bulge (Carney \etal\ 1990; Wyse \& Gilmore
1992). 

There is little scatter in the trend of element ratios, such as
[Mg/Fe] against [Fe/H], for the bulk of the stellar halo (e.g.~Nissen
\etal\ 1994), consistent with enrichment by stars with a fixed
massive-star IMF, and furthermore,  one close to that observed in star-forming
regions locally today (see Wyse 1998 for a review).  The lack of scatter 
also implies rather efficient mixing, and one might expect
to see some scatter at the lowest levels of enrichment, when very few
supernovae have contributed (cf.~Audouze \& Silk 1995; Tsujimoto, 
Shigeyama \& Yoshii 1999), and this is indeed observed (McWilliam 
et al.~1995; Ryan, Norris \& Beers, 1996;  Ryan this volume). 

In this picture of star-forming fragments in the halo, there is the
possibility that a few fragments had a deep enough potential-well to
sustain star formation and self-enrich with the products of Type Ia
supernovae.  Further, with asynchronous onsets of star formation in
different regions, there could be enrichment of a given `fragment' by
a Type Ia supernova whose progenitors were formed in a
different fragment.  Thus one might expect to see at least some stars
in the halo now with values of the element ratios reflecting `extra'
iron from Type Ia supernovae (note that the fact that this is not
generally observed for halo stars is further motivation for the prompt
removal of gas from the halo to the bulge; Wyse \& Gilmore 1992).  Indeed, 
a subset of halo stars, apparently biased towards the metal-rich halo, 
i.e. [Fe/H]$ > -1$~dex (Nissen \& Schuster 1997) and/or extremely high-energy orbits (Carney \etal\ 1997; King 1997; Ivans \etal\ this volume)  
have now been observed to have low values of
[Mg/Fe], close to or below the solar value, consistent with
pre-enrichment by a combination of Type I and Type II supernovae.

\begin{figure}[!htb]
\hskip 0.5in
\psfig{file=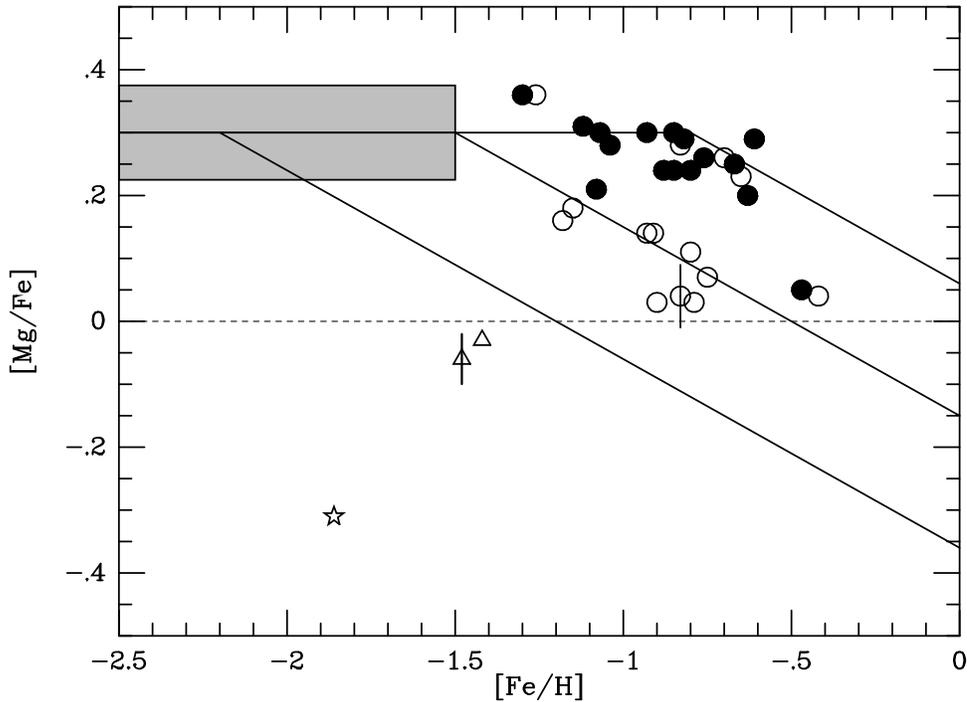,width=5.5in}
\vskip -.6truecm
\caption{Element ratio [Mg/Fe] against iron abundance for the
kinematically-defined disk stars (open circles) and the kinematically-defined halo stars (filled
circles) from Nissen \& Schuster (1997), together with  the metal-poor 
anomalous halo stars from King (1997;
triangles) and from Carney \etal\ (1997; star symbol); uncertainties in [Mg/Fe] as indicated.  The lines drawn
to guide the eye are schematic trends of [Mg/Fe] against [Fe/H] for
closed, self-enriching systems of invariant IMF and a range of star formation
rates; a higher star formation rate leads to a higher value of the
iron abundance at the onset of Type Ia supernovae, which produces a
downturn in this plot.  The shaded region indicates the locus of normal halo stars. 
There are three main points to this figure (based on figure~1 of 
Gilmore \& Wyse 1998), 
the
first being that the value of the Type II [Mg/Fe] `plateau' is the same for
both the halo and disk stars, the second being that the 
vast majority of `metal-rich' halo stars
have  lower values of [Mg/Fe] than this canonical `plateau' value, 
consistent with
having `extra' iron from Type Ia supernovae, and the third 
being that the typical elemental abundances of disk and  
halo stars of the same iron abundance are different, supporting a 
disk/halo discontinuity.}

\end{figure}

Fig.~5 is taken from Gilmore \& Wyse (1998) and based on the data of
Nissen \& Schuster (1997) for their survey of disk and halo stars of
similar metallicities, where `disk' and `halo' are defined
kinematically in terms of orbital rotational velocity.  This is one of
the few datasets where both disk and halo stars have been analysed
together, allowing a direct comparison of their elemental abundances.
As can be seen, there are both disk and halo stars with enhanced
magnesium, [Mg/Fe]$ \sim +0.3$, consistent with being enriched by
massive stars with the same, invariant IMF.  The halo stars and disk
stars with low values of [Mg/Fe] are as expected for some enrichment
from Type Ia supernovae. Note that in Fig.~5 the halo stars lie along
the locus for a lower star-formation rate than do the disk stars, but
this may be a manifestation of the fact that gas outflows, as invoked
above to have operated in halo star-forming regions, reduce the
efficiency of star formation and mimic a lower value.

The lower values of the $\alpha$-elements in these low-metallicity
stars (remember that the higher metallicity part of the halo is still
well below solar metallicity) are as predicted for the metal-poor
stars in dwarf companion galaxies to the Milky Way (Gilmore \& Wyse
1991). This, combined with the fact that the 
serendipitously-identified metal-poor halo stars with anomalously low
values of the ratio of magnesium-to-iron are on retrograde orbits, 
led to the speculation that these stars may have been captured
(accreted) from external satellite galaxies (Carney \etal\ 1997; King
1997).  However, there is no tendency for the low [Mg/Fe] stars in the
Nissen \& Schuster sample to be on retrograde orbits.  All the
`low-alpha-elements' stars are, however, on very high-energy radial
orbits, with apoGalacticon greater than 15~kpc, and periGalacticon
less than 1~kpc.  These are very unlikely orbits for stars accreted
from satellite galaxies, a term that implies a separate identity 
for a significant time.  In models which invoke fragmentation
within a gaseous proto-halo, fragments which probe the denser inner
Galaxy are naturally themselves more dense (e.g.~Fall \& Rees 1985),
and likely to have deeper local potential wells, and be able to
self-enrich longer.  Thus a trend that the halo stars with evidence
for enrichment by Type Ia supernova be on orbits of low periGalacticon
(but not necessarily high apoGalacticon) 
may be understood, without any need to appeal to `accretion'.
Further, Stephens (1999) has analysed a sample of halo stars selected
to be on extreme orbits, and interprets the elemental abundance
patterns as being no different from those of the rest of the halo.
However, there is clearly a need for a large, uniformly-analysed
sample over the entire range of kinematics and metallicity of the
stellar halo.

A further point evident from Fig.~5 is that the typical elemental
abundances of disk and halo stars of the same iron abundance are
different, supporting a disk/halo discontinuity in chemical
enrichment, and consistent with the different specific angular
momentum content of disk and halo (Wyse \& Gilmore 1992) -- the local
halo did {\it not\/} pre-enrich the local disk.  These metal-rich halo
stars are a small fraction of the locally-defined stellar halo.
However, it must be noted that the {\it global\/} metallicity
distribution of the stellar halo is poorly defined (in particular for
the inner halo), as are the wings of the metallicity distribution
function for more locally-defined samples.  The overlap between
stellar halo and (thick) disk at metallicities around [Fe/H]$ = -1$ is
a particular focus of our ongoing 2dF project (Gilmore, Wyse, Norris
\& Freeman).

\subsection{Kinematics \& Age}

Moving groups of stars in the halo have been searched for by many
groups, with limited success in that the signatures are usually of low
statistical weight (e.g.~Arnold \& Gilmore 1992; Majewski, Munn \&
Hawley 1996; Helmi \& White 1999; Helmi, White, de Zeeuw \& Zhao
1999).  A complication in the interpretation of moving groups is that
all substructure in the Galaxy is subjected to tidal effects.  For
example, the present system of globular clusters may well be a mere
shadow of the initial retinue (e.g.~Gnedin \& Ostriker 1997) and one
expects e.g.~tidal arms and streamers from the surviving globular
clusters (see Grillmair, Freeman, Irwin \& Quinn, 1995; Meylan, this
volume).
 
The unique substructure that has been found by virtue of its location
in position--radial velocity phase space is the Sagittarius Dwarf
Spheroidal galaxy (Ibata, Gilmore \& Irwin 1994, 1995).  This
satellite companion to the Milky Way was noticed by its discoverers,
in the course of their survey of stars in the bulge of the Milky Way,
as a distinct set of stars with `anomalous' kinematics to be actually
members of the bulge. The stars with these very well-defined, but
anomalous, kinematics were localised in a subset of their
lines-of-sight, and could be identified with a feature (the red clump
of helium-burning stars) in the colour-magnitude diagram for all stars
in those fields.  The center of the Sagittarius dwarf is located some
24~kpc from the Sun, on the other side of the Galactic centre. 
The preliminary proper motion and orbit
derived by Ibata \etal\ (1997) for this dwarf galaxy imply that it 
has a radial period of less than 1~Gyr, and
a periGalacticon of only $\sim 12$~kpc.  
Without a significant amount of dark matter to bind it,
the Sagittarius dSph would be unable to survive more than a couple of
such close periGalactic passages (e.g.~Velazquez \& White 1995), but
its dominant stellar population has an age of $\sim 10$~Gyr.  Either
the orbital parameters have been recently changed (e.g.~Zhao 1998), or
the dwarf is more robust than it looks (Ibata \etal\ 1997). 

The more diffuse outer regions of the Sagittarius dSph are more
susceptable to tidal stripping, and intriguing observations of stars
with kinematics such that they plausibly could have been removed from
the dwarf galaxy on a previous periGalactic passage have been reported
by Majewski \etal\ (1999), and analysed in Johnston \etal\
(1999). Further, this interpretation is consistent with the
photometric detection of member stars some 30$^\circ$ from the centre of the
Sagittarius dwarf by Mateo, Olszewski \& Morrison (1998).

A few of the globular clusters of the Milky Way have clearly younger
ages than the vast majority of globular clusters and have been
suggested as candidates for being accreted from companion galaxies
(e.g.~ Fusi Pecci \etal\ 1995). Indeed it is now recognised that this
group includes actual members of the Sagittarius dwarf's retinue of
clusters (e.g.~Ibata \etal\ 1995; Da Costa \& Armandroff 1995), and
should not be included in the census of Milky Way clusters.  The
globular cluster M54 is situated at the centre of the Sagittarius
dwarf, no doubt partly motivating the interpretation of the globular cluster
$\omega$~Cen as the nucleus of an accreted dwarf galaxy (Majewski
\etal\ this volume; Wallerstein \& Hughes, this volume).  The very
different stellar populations in a typical companion galaxy and in the
stellar halo provide strong constraints on general accretion and
disruption of satellite galaxies as a means to form the stellar halo
(Unavane, Wyse \& Gilmore 1996; Gilmore this volume).  
Indeed significant (greater than
$\sim 10$\% of the stellar halo) accretion cannot have occurred from 
typical dwarf galaxies, with their large range of stellar ages, subsequent
to the formation of these intermediate-age stars, and is thus
restricted to $\geq 8$~Gyr ago.

Several analyses of chemistry and kinematics for samples of halo stars
have found evidence for differences between the `far' halo and the
`inner' halo e.g.~ Majewski (1992), Norris (1994), Carney \etal\
(1996), Layden (1998), Carney (this volume). 
This would plausibly reflect differences in the
dominant physical mechanisms at formation.  However, for any
reasonable halo profile the fractional mass of the `far' halo is small
(see e.g.~Fig.~1 of Unavane, Wyse \& Gilmore 1996).  Further, Carney
(private communication) has found intriguing differences in binary fraction
between the `far' halo and the `inner' halo, in that the `far' halo
has a significantly lower fraction.  Mass transfer in close binary
systems could certainly affect surface elemental abundances, and the
`kick' from disruption could perhaps produce extreme kinematics.
Clearly more work is warranted.

\subsection{Faint Stellar Luminosity Function and IMF}

As discussed briefly above in Sect.~32.1, the elemental abundances for
the bulk of the stellar halo and the disk suggest that the massive
star IMF was, and is, invariant.  Low-mass stars, those with
main-sequence lifetimes that are of order the age of the Universe,
provide more direct constraints on the IMF when they formed.  Star
counts in systems with simple star-formation histories are
particularly straightforward to interpret, and those in `old' systems
allow one to determine the low-mass stellar IMF at large look-back
times and thus at high redshift.  The dwarf spheroidal satellite
galaxies of the Milky Way are now accessible for this experiment using
the Hubble Space Telescope.  These galaxies are particularly
interesting since their internal kinematics suggest that they are
among the most dark-matter-dominated systems known (reviewed by Mateo
1998), and this dark matter must be cold to form structures on such
small scales (e.g.~Tremaine \& Gunn 1978; Gerhard \& Spergel 1992),
but at least the gas-rich dwarfs for which rotation curves can be
measured do not fit the predictions of non-baryonic CDM (e.g.~Moore
1984).  Could the dark matter be cold since it is baryonic and
radiated away binding energy?  Might the dark matter then be
associated with faint stars?

The Ursa Minor dwarf spheroidal galaxy (dSph) is  suitable for study, being  
relatively nearby (distance $\sim 70$kpc), 
and, unusually for a dwarf spheroidal galaxy,  
having a stellar population with narrow distributions of age
and of metallicity (e.g.~Hernandez, Gilmore \& Valls-Gabaud 1999), 
remarkably similar to that of a classical halo globular
cluster such as M92 or M15, i.e. old and metal-poor ([Fe/H] $ \sim
-2.2$~dex).  The integrated luminosity of the Ursa Minor dSph ($L_V
\sim 3 \times 10^5 L_\odot$) is also similar to that of a globular
cluster. However, the central surface brightness of the Ursa Minor
dSph is only 25.5 V-mag/sq arcsec, corresponding to a central
luminosity density of 0.006~$L_\odot$pc$^{-3}$, many orders of
magnitude lower than that of a typical globular cluster.  Further,
again in contrast to globular clusters, its internal dynamics are
dominated by dark matter, with $(M/L)_V \sim 80$, based on the
relatively high value of its internal stellar velocity dispersion
(Hargreaves {\it et al.} 1994; see review of Mateo 1998).

We obtained deep imaging data with the Hubble Space Telescope, using
WFPC2 (V-606 \& I-814), STIS (LP optical filter) and NICMOS (H-band),
in a field close to the center of the Ursa Minor dSph (program
GO~7419: PI~Wyse, Co-Is Gilmore, Tanvir, Gallagher, Smecker-Hane,
Feltzing \& Houdashelt).  As shown in Fig.~6 (from Feltzing, Gilmore
\& Wyse 1999), the faint optical stellar luminosity function of the
Ursa Minor dSph is also remarkably similar to that of M92, down to our
limiting apparent magnitude with WFPC2 data, which corresponds to
around four-tenths of a solar mass.  The M92 data (Piotto, Cool \&
King 1997) should be a reliable estimate of the global initial
luminosity (and mass) function in this cluster, being obtained at
intermediate radius within the globular cluster, minimising internal
dynamical effects, and the cluster itself is on an orbit that
minimises external tidal effects. The similarity of age and
metallicity between these two systems means that the comparison of the
main sequence faint luminosity functions is effectively a comparison
of stellar initial mass functions.  And as can be seen in the figure,
these two luminosity functions are remarkably similar, down to our
completeness limit corresponding to around 0.4~$M_\odot$. Thus two
systems that differ in mass to light ratio by a factor of roughly 50,
and in stellar surface density by orders of magnitude, formed stars
with the same initial mass function.  A consistent result, but for a
significantly less-deep luminosity function reaching to 0.6$M_\odot$,
was obtained for the Draco dSph by Grillmair \etal\ (1998).

The apparent insensitivity of the stellar IMF to any
parameter that physical intuition tells one should be important is
remarkable (see papers in Gilmore \& Howell, 1998).  However, it
allows a reliable simplifying assumption -- an invariant IMF -- to be
made when modelling the evolution of galaxies.

\begin{figure}
\psfig{file=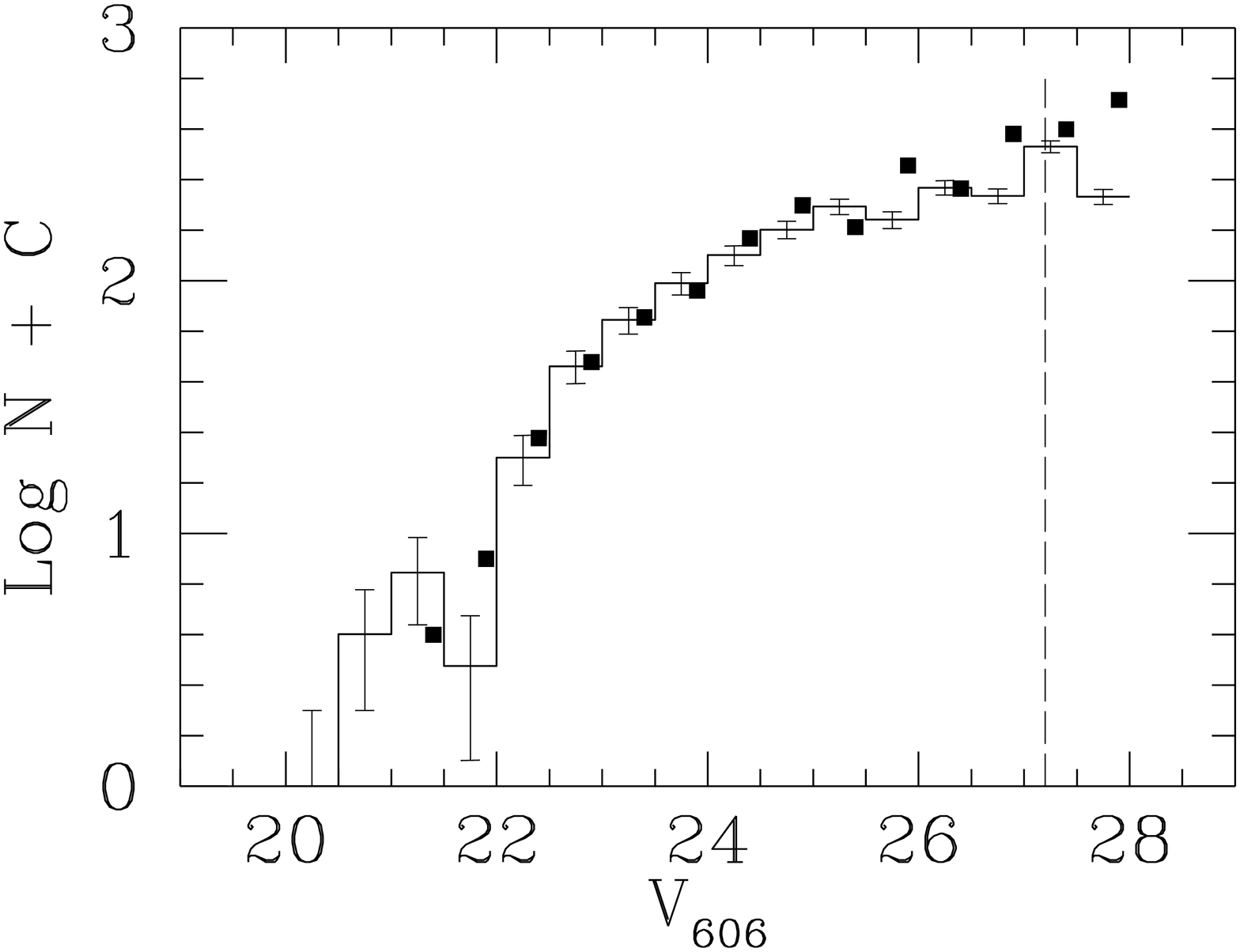,width=3in,angle=270}
\vskip -6cm
\hskip 3.5in
\psfig{file=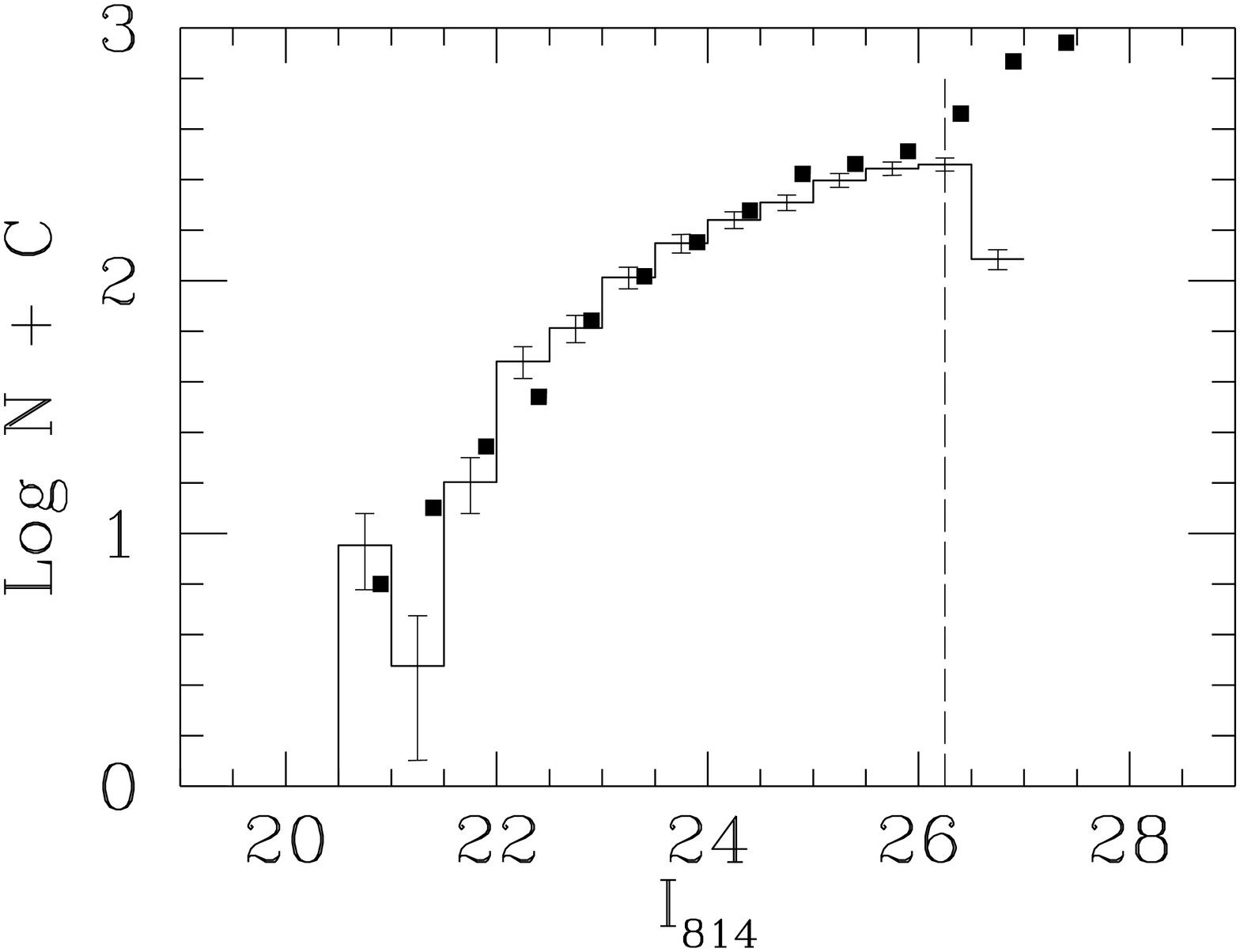,width=3in,angle=270}

\vskip -0.25truecm
\caption{Comparison between the completeness-corrected Ursa Minor
luminosity functions (histograms; 
50\% completeness indicated by the vertical dashed line) 
in the V-band (left panel) and the I-band (right panel)  and the same 
for M92 (points, moved to same distance as the Ursa Minor dSph; 
from Piotto {\it et al.} 1997). The luminosity functions for the 
globular cluster and the dwarf spheroidal galaxy are indistinguishable.}
\end{figure}

The identity of the dark matter in dSph galaxies remains a puzzle; we
are obtaining deeper data and should be able to push the stellar
luminosity function somewhat fainter.  However, low-mass stars do not
look likely candidates.

\section{Summary and Conclusions}

`Scenario' rather than `model' is appropriate for the title of this
paper, since we do not yet have a clear understanding of the
mechanisms by which disk galaxies form and evolve.  However, it is
clear that substructure plays an important role in galaxy formation
and evolution. For example, the thick disk is plausibly a remnant of
the last significant merger event in the history of the Milky Way; the
remnant satellite may have been detected. That this last significant
merger was a long time ago is apparently in conflict with flat
CDM-dominated models.  Further, while accretion of stars from fragile
satellite galaxies may make a significant contribution to the outer
halo, this is not the case for the bulk of the halo; late accretion in
particular is ruled out by the uniform old age of the bulk of the halo
stars.  A simplifying factor for models of galaxy evolution is that
the stellar IMF is apparently invariant.

%
\section*{Acknowledgements}
I would like to thank the organisers for their financial help  to
enable me to participate in the meeting, and for arranging such a
stimulating and enjoyable conference.  I acknowledge partial support from  
NASA ATP program grant NAG 5-3928. 
Support for my work on the Ursa
Minor dSph luminosity function was provided by NASA through
grant 
GO-07419.01-96A from the Space Telescope Science Institute, which is operated
by AURA, under NASA contract NAS5-26555. 
%
%
 
\beginrefer

\refer Armandroff, T., 1989, AJ 97, 375

\refer Arnold, R. \& Gilmore, G., 1992, MNRAS 257, 225

\refer Audouze, J. \& Silk, J. 1995, ApJL 451, L49

\refer Bahcall, N., Ostriker, J., Perlmutter, S. \& Steinhardt, P., 1999, Science, 284, 1481

\refer Baugh, C.M., Cole, S., Frenk, C.S. \& Lacey, C.G., 1998, ApJ 498, 504


\refer Blitz, L., Spergel, D.N., Teuben, P.J., Hartmann, D. \& Burton,
W.B., 1999, ApJ 514, 818

\refer Bonifacio, P., Pasquini, L., Molaro, P. \&  Marconi, G.  1999, 
in `Galaxy Evolution: Connecting the Distant Universe with the Local 
Fossil Record', Obs de Meudon, 1998.

\refer Burkert, A. \& Silk, J., 1997, ApJL 488, L55 

\refer Burkert, A. \& Silk, J., 1999, in `Dark Matter in Astro and Particle 
Physics', ed. H.V. Klapdor-Kleingrothaus, in press

\refer Burkert, A. \& Yoshii, Y., 1996, MNRAS 282, 1349

\refer Burkert, A., Truran, J. \& Hensler, G., 1992, ApJ 391, 651

\refer Burstein, D., 1979, ApJ 234, 829


\refer Carney, B., Latham, D.  \& Laird, J., 1989, AJ 97, 423 

\refer Carney, B., Latham, D.  \& Laird, J., 1990, AJ 99, 572

\refer Carney, B., Latham, D., Laird, J. \& Aguilar, L., 1994, AJ 107, 2240

\refer Carney, B., Laird, J., Latham, D.  \&  Aguilar, L., 1996, AJ 112, 668

\refer Carney, B., Wright, J., Sneden, C.,  Laird, J., Aguilar, L. \&  
Latham, D., 1997, AJ 114, 363

\refer Da Costa, G., 1999, in `The Galactic Halo', Third Stromlo Symposium, 
ASP Conference Series vol 165, eds B.~Gibson, T.~Axelrod \& M.~Putman (ASP, San Francisco) p153

\refer Da Costa, G. \& Armandroff, T., 1995, AJ 109, 2533

\refer Dalcanton, J.J., Spergel, D.N., \& Summers, F.J., 1997, ApJ 482, 659.   

\refer de Freitas Pacheo, J., Barbuy, B. \& Idiart, T., 1998, A\&A 332, 19

\refer Edvardsson, B., Andersen, J., Gustafsson, B., Lambert, D.L., 
Nissen, P.E. \& Tomkin, J., 1993, A\&A 275, 101

\refer Eggen, O., Lynden-Bell, D. \& Sandage, A., 1962, ApJ 136, 748

\refer  Fall, S.M. \& Efstathiou, G., 1980, MNRAS 193, 189  

\refer Fall, S.M. \& Rees, M.J., 1985, ApJ 298, 18
 
\refer Feltzing, S. \& Gilmore, G., 1999, A\&A in press

\refer Feltzing, S., Gilmore \& Wyse, R.F.G. 1999, ApJL 516, L17

\refer Freeman, K. 1993, in `Galaxy Evolution: The Milky Way
Perspective', ASP Conference series vol 49, ed.~S.~Majewski (ASP, San
Francisco) p125

\refer Fry, A., Morrison, H., Harding, P. \& Boroson, T., 1999, AJ 118, 1209

\refer Fuchs, B., Jahreiss, H. \& Wielen, R., 1999, in `Galaxy
Evolution: Connecting the Distant Universe with the Local Fossil
Record', Obs de Meudon, 1998.

\refer Fuhrmann, K., 1998, A\&A 338, 161

\refer Fusi Pecci, F., Bellazzini, M., Cacciari, C. \&  Ferraro, F.R., 1995, AJ 110, 1664

\refer Gerhard, O.E. \& Spergel, D.N., 1992, ApJL 389, L9

\refer Gilmore, G., \& Howell, D., 1998, eds `The  Stellar IMF', 
ASP Conference series vol 142,  (ASP, San Francisco)

\refer Gilmore, G. \& Reid, I.N., 1983, MNRAS 202, 1025

\refer Gilmore, G. \& Wyse, R.F.G., 1985, AJ 90, 2015

\refer Gilmore, G. \& Wyse, R.F.G., 1991, ApJ 367, L55

\refer Gilmore, G. \& Wyse, R.F.G., 1998, AJ 116, 748

\refer Gilmore, G., Wyse, R.F.G. \& Jones, J.B., 1995, AJ 109, 1095
 
\refer Gilmore, G., Wyse, R.F.G. \& Kuijken, K., 1989, ARAA 27, 555

\refer Gnedin, O.Y. \& Ostriker, J.P., 1997, ApJ 474, 223

\refer Grillmair, C., Freeman, K.C., Irwin, M. \& Quinn, P.J., 1995, AJ 109, 2553

\refer Grillmair, C., \etal\ 1998, AJ 115, 144

\refer Hargreaves, J., Gilmore, G., Irwin, M. \& Carter, D., 1994, MNRAS 
271, 693

\refer Hartwick, F.D.A., 1976, ApJ 209, 418

\refer Helmi, A. \& White, S.D.M., 1999, MNRAS 307, 495

\refer Helmi, A., White, S.D.M., de Zeeuw, P.T. \& Zhao, H.-S. 
1999, Nat 402, 53 

\refer Hernandez, X., Gilmore. G. \& Valls-Gabaud, D., 1999, MNRAS in press

\refer Hernquist, L. \& Mihos, J.C., 1995, ApJ 448, 41

\refer Huang, S. \& Carlberg, R.,  1997, ApJ 480, 503

\refer Ibata, R. \& Gilmore, G., 1995, MNRAS 275, 605

\refer Ibata, R., Gilmore, G. \& Irwin, M., 1994, Nat 370, 194

\refer Ibata, R. Gilmore, G. \& Irwin, M., 1995, MNRAS 277, 781

\refer Ibata, R.,  Wyse, R.F.G., Gilmore, G., Irwin, M.J. \& Suntzeff, N.B., 1997, AJ 113, 634

\refer Johnston, K.V., 1998, ApJ 495, 297

\refer Johnston, K.V., Majewski, S., Siegel, M., Reid, I.N. \& Kunkel, W.E., 1999, AJ in press

\refer Kauffmann, G., Colberg, J.M., Diaferio, A. \& White, S.D.M., 1999, MNRAS 303, 188

\refer King, J., 1997, AJ 113, 2302

\refer  Klypin, A., Kravtsov, A.V.,  Valenzuela, O. \&  Prada, F., 1999, ApJ 522, 82

\refer Lacey, C.G., 1991, in `Dynamics of Disc Galaxies', ed B.~Sundelius (Goteborg University, Sweden) p257

\refer Lacey, C.G. \& Cole, S., 1993, MNRAS 262, 627

\refer Layden, A.C., 1998, in `Galactic Halos', ASP Conference series vol.~136 ed D.~Zaritsky (ASP, San Francisco) p14

\refer McWilliam, A., Preston, G., Sneden, C. \& Searle, L. 1995, AJ 109, 2757 

\refer Majewski, S.R. 1992, ApJS 78, 87

\refer Majewski, S.R., 1993, ARAA 31, 575

\refer Majewski, S.R, Munn, J. \& Hawley, S., 1996, ApJL 459, L73

\refer  Majewski, S., Siegel, M., Kunkel, W.E., Reid, I.N., Johnston, K.V., Thomson, I., Landolt, A \& Palma, C. 1999, AJ in press

\refer Mateo, M., 1998, ARAA 36, 435

\refer Mateo, M., Olszewski, E. \& Morrison, H., 1998, ApJL 508, L55
 
\refer Mendez, R., Platais, I., Girard, T., Kozhurina-Platais, V. \& 
Altena, W., 1999, AJ in press

\refer Minniti, D., 1996, ApJ 459, 175

\refer Mo, H.J., Mao, S.  \& White, S.D.M., 1998, MNRAS 295, 319 

\refer  Moore, B., 1994, Nat 370, 629

\refer Moore, B. Ghigna, S.,  Governato, F.,  Lake, G., Quinn, T., Stadel, 
J. \& Tozzi, P., 1999a, ApJL 524, L19

\refer Moore, B., Quinn, T., Governato, F., Stadel, J. \& Lake, G., 1999b, MNRAS submitted

\refer Morrison, H., 1993, AJ 106, 578

\refer Navarro, J.F. \& Steinmetz, M., 1997, ApJ 478, 13

\refer Navarro, J.F., Frenk, C.S. \& White, S.D.M., 1997, ApJ 490, 493  

\refer Nissen, P. \& Schuster, W., 1997, A\&A 326, 751

\refer Nissen, P., Gustafsson, B., Edvardsson, B. \& Gilmore, G., 1994, A\&A  
285, 440

\refer Norris, J., 1994, ApJ 431, 645 

\refer Norris, J., 1999, in `The Galactic Halo', Third Stromlo Symposium, 
ASP Conference Series vol 165, eds B.~Gibson, T.~Axelrod \& M.~Putman (ASP, San Francisco) p213

\refer Ortolani, S., Renzini, A., Gilmozzi, R., Marconi, G., Barbuy,
B., Bica, E. \& Rich, R.M., 1995, Nat 377, 701

\refer Ostriker, J.P., 1990, in `Evolution of the Universe of Galaxies', 
ASP Conference Proceedings Vol. 10, eds. Kron, R.G., p25

\refer Ostriker, J.P. \& Lacey, C., 1985, MNRAS 299, 633

\refer Pearce, F. \etal\ (the VIRGO Consortium), 1999, ApJL 521, 99L

\refer Phleps, S., Meisenheimer, K., Fuchs, B., Wolf, C. \& Jahreiss,
H., 1999, in `Galaxy Evolution: Connecting the Distant Universe with
the Local Fossil Record', Obs de Meudon, 1998.

\refer Piotto, G., Cool, A. \& King, I.,  1997, AJ, 113, 1345

\refer Putman, M., {\it et al.}, 1998, Nat 394, 752
 
\refer Rocha-Pinto, H. \& Maciel, W., 1997, MNRAS 289, 882

\refer Ryan, S., Norris, J. \& Beers, T., 1996, ApJ 471, 254

\refer Sanchez-Salcedo, F.J., 1999, MNRAS 303, 755

\refer Shaw, M. \& Gilmore, G., 1990, MNRAS 242, 59

\refer Smecker, T. \& Wyse, R.F.G., 1991, ApJ 372, 448

\refer Spitzer, L. \& Schwartzschild, M., 1953, ApJ 118, 106 

\refer Stephens, A., 1999, AJ 117, 1771

\refer Tegmark, M., Silk, J., Rees, M.J., Blanchard, A., Abel, T. \&
Palla, F., 1997, ApJL 474, 1L

\refer Toth, G. \& Ostriker, J.P., 1992, ApJ 389, 5

\refer Tremaine, S. 1993, in `Back to the Galaxy', eds S.~Holt \& F.~Verter (AIP, New York) p599

\refer Tremaine, S. \& Gunn, J., 1978, Phys Rev Lett 42, 407

\refer Tsikoudi, V., 1979, ApJ 234, 842

\refer Tsujimoto, T., Shigeyama, T. \& Yoshii, Y., 1999, ApJL 519, L63

\refer Unavane, M., Wyse, R.F.G. \& Gilmore, G., 1996, MNRAS 278, 727

\refer VandenBerg, D., 1985, ApJS 58, 711

\refer VandenBerg, D. \& Bell, R., 1985, ApJS 58, 561

\refer Velazquez, H. \& White, S.D.M., 1995, MNRAS 275, L23

\refer Velazquez, V. \& White, S.D.M., 1999, MNRAS 304, 254 

\refer  Walker, I., Mihos, J.C. \& Hernquist, L., 1996, ApJ 460, 121

\refer Weil, M., Eke, V. \& Efstathiou, G., 1998, MNRAS 300, 773

\refer White, S.D.M. \& Rees, M.J., 1978, MNRAS 183, 341

\refer Wielen, R., 1977, A\&A 60, 263

\refer Wyse, R.F.G., 1998, in `The Stellar IMF', ASP Conference series 
vol 142, eds G.~Gilmore \& D.~Howell, (ASP, San Francisco) p89

\refer Wyse, R.F.G. \& Gilmore, G., 1990, in `Chemical \& Dynamical
Evolution of Galaxies', eds F.~Ferrini, J.~Franco \& F.~Matteucci (ETS
Editrice, Pisa, Italy) p19

\refer Wyse, R.F.G. \& Gilmore, G., 1986, AJ 92, 1215

\refer Wyse, R.F.G. \& Gilmore, G., 1992, AJ 104, 114 

\refer  Wyse, R.F.G. \& Gilmore, G., 1995, AJ 110, 2771

\refer Wyse, R.F.G., Gilmore, G. \& Franx, M., 1997, ARAA 35, 637

\refer Yungelson, L. \& Livio, M. 1998, ApJ 497, 168

\refer Zhang, B. \& Wyse, R.F.G., 2000, MNRAS in press

\refer Zhao, H.-S. 1998, ApJL 500, L149

\endrefer           
\end{document}